# Using Computer Vision Techniques for Moving Poster Design

Rebelo, Sérgio[1]; Martins, Pedro[1]; Bicker, João[1]; Machado, Penousal[1].
srebelo@dei.uc.pt; pjmm@dei.uc.pt; bicker@dei.uc.pt; machado@dei.uc.pt
[1] CISUC, Department of Informatic Engineering, University of Coimbra,
3030 Coimbra, Portugal

**Abstract:**
Graphic Design encompasses a wide range of activities from the design of traditional print media (e.g., books and posters) to site specific (e.g., signage systems) and electronic media (e.g., interfaces). Its practice always explores the new possibilities of information and communication technologies. Therefore, interactivity and participation have become key features in the design process. Even in traditional print media, graphic designers are trying to enhance user experience and exploring new interaction models.

Moving posters are an example of this. This type of posters combine the specific features of motion and print worlds in order to produce attractive forms of communication that explore and exploit the potential of digital screens. In our opinion, the next step towards the integration of moving posters with the surroundings, where they operate, is incorporating data from the environment, which also enables the seamless participation of the audience. As such, the adoption of computer vision techniques for moving poster design becomes a natural approach.

Following this line of thought, we present a system wherein computer vision techniques are used to shape a moving poster. Although it is still a work in progress, the system is already able to sense the surrounding physical environment and translate the collected data into graphical information. The data is gathered from the environment in two ways: (1) directly using motion tracking; and (2) indirectly via contextual ambient data. In this sense, each user interaction with the system results in a different experience and in a unique poster design.

## 1. Introduction

Technology always shaped and was shaped by every aspect of the society. In this sense, Graphic Design (GD) — the discipline responsible for organising visual communication in our society (Frascara, 1988) —, has always been keeping a symbiotic development with the technological improvements (Cooper, 1989). The recent technological advances, since of the last quarter of the twentieth century, are no exception. Nowadays' practice of GD is located between the traditional approaches and the use of the new information and communication technologies (Neves, 2011). Besides, today's audience is changing and no longer contents to simply digest messages (Armstrong & Stojmirovic, 2011). The contemporary public increasingly approaches design's artefacts with the expectation of interaction. This started to force the mixing of the traditional print and location-specific media with digital electronic media. Posters are no exception. The digital screens, which have infiltrated all kind of spaces, set up a new design challenge, allowing the inclusion of video and animated GIF (Benyon, 2016). However, as in other moments in history, a new generation of designers are embracing these technologies, in order to combine experiences and to create a novel form of communication — *The Moving Poster*.

This new medium of communication, which is not a poster, in its traditional sense, does not present itself as an interactive application (e. g., websites or digital application). It inherited the traditional vertical poster format and its composition rules, but often has a digital version with different temporal states (e.g. (Pfäffli, 2014), (Studio Feixen & Giger, 2017) or (Creative Review, 2017)). According to Brechbühl (Benyon, 2016) it still is a poster, but "the animation part is more like a fifth colour, or a special print technique."

Typically, the moving posters are designed based on a set of temporal states, displayed to the user in a sequential way. These posters border on storytelling, but still only communicate the same information (Benyon, 2016). Although in some artefacts, the interactive and generative principles are starting to be taken into consideration (e.g. (LUST, 2014)). According to the *Screen Design* principles point of view, its development and implementation are still not well achieved (see (Macklin, 2008)). Concepts as ergonomics/human factors, human-computer interaction (HCI) and Interaction Design (IxD) may integrate more levels of information to fully engage the viewer. In the coming periods, this kind of posters will become mainstream and will begin to respond to their environment using information gathered by input devices (as such cameras, RFID tags, or audio devices) (Benyon, 2016). Thereby, digital technologies will enable dialogues between the artefact and the viewers and will promote the seamless participation of the user and the serendipity of data derived from a process-oriented GD (Armstrong & Stojmirovic, 2011).

Following this line of thought, we consider the adoption of Computer Vision (CV) techniques in this scenario as a natural step towards the moving posters' strategic definition. In this sense, we develop experiments in the design of a moving poster wherein CV techniques are used to define the poster's shape. However, contrary to current mainstream approaches, the poster is shaped through live input data gathered by user tracking. In this way, the user can directly manipulate the graphical elements in the composition, and at the same time see the result of the manipulation. Although it is still a work in progress, in these first experiments the data from the user and from the surrounding environment already shapes the elements of the poster.

The remainder of this paper is organised as follows: Section 2 briefly presents the background of the field; Section 3 thoroughly describes how the poster is designed; Section 4 describes the poster element's movements; and finally, conclusions and future work are presented in Section 5.

## 2. Background

Desktop Publishing (DTP) revolution led Graphic Design (GD) to be the first profession reshaped by digital technologies (Blauvelt, 2011). Contrary to the strong resistance of several designers to DTP technologies, in the turn of the millennium, computer have already adopted as GD primary tool (Meggs & Purvis., 2011; Blauvelt, 2011). Besides, desktop publishing presented new ways of displaying information and approaching grids, changing permanently the methods of design (Licko & Vanderlands, 1989; Meggs & Purvis., 2011). Thereby, GD was transformed into a more interdisciplinary and ubiquitous field. These characteristics increased whenever digital technologies embed a bit further into our everyday lives, creating more "spaces" to be designed (Lupton, 2006).

While graphic designers embrace and use recent digital technologies, they do not fully understand them (Neves, 2011). Like in the 1990s, nowadays the computer continues to be viewed by several designers as another tool — just like a different type of pencil (Maeda, 2004). Although some designers, especially from the newest generations, understand the possibilities of using the computer as a medium of expression, they lack the technical knowledge, particularly in computer programming and Interaction Design (IxD) (Benyon, 2016). Limitations that restrain the outreach of graphic design projects and, so, hinder an all-encompassing exploration of the digital media.

The emerging interest in the use of data to produce aesthetic artefacts proves this point. Currently, creative disciplines are seduced by the fusion of different media and by concepts as Data-Driven Design, Generative Design, Parametric Design, or Information Design (Laranjo, 2017). As mentioned above, moving posters are a natural example of this change in basic assumptions of GD.

In recent times, moving posters have become popular in GD's landscape. However, a discussion about what constitutes a good moving poster, and what is its medium's boundaries was started (Benyon, 2016; Thiriet, 2017). In this sense, specialized competitions appeared, such as the *Typomanina Video Poster Contest* (see (Typomania, 2017)) or the *Weltformat Poster Festival* (see (Weltform, 2017)). Other festivals also started to accept applications of this medium, in their poster exhibitions and contests. For instance, the *Graphic Design Festival Scotland* (see (Graphic Design Festival Scotland, 2017)).

The first steps to the definition of the medium were taken by Josh Schaub, with the creation of the web repository *The Moving Poster* (Schaub, n. d.). This database collects examples of 'moving posters' to present the possibilities and the limitations of the work in this medium. (Benyon, 2016). As the main objective, the inventory seeks to answer questions such as: (1) What are the techniques and methods of narration?; (2) Where does the poster ends and where does a film begin?; and (3) What a poster actually is and how this medium will continue to develop in the future? (Schaub, n. d.). Currently, the inventory has a total of fifty-three posters.

Additionally, Schaub (Schaub, n. d.) also developed a taxonomy for classifying the works he is collating. His classification considers that posters can be divided into a spectrum from that goes from static to filmic designs. The static posters are those whereby no structural alterations take place, there are not position changes and the animation effects are predominantly static (e.g. altering colour surfaces or single visuals). In the middle of this spectrum are the dynamic posters. A dynamic poster is designed with moving parts. This segment can be broken into innumerable formats and categories. Finally, the filmic posters are those whose the movement is exclusively focused on certain filmic scenes (e.g. a moving background).

In the context of this paper, we are not interested in the type of animation but in seeking artefacts that provide interactions that are triggered by the public. These posters can be in the whole spectrum defined by (Schaub, n. d.). In this sense, we propose an alternative classification to 'moving posters' in two subsets: (1) animated, and (2) interactive.

The animated posters present the information in a sequential way (from a Start key frame to an End key frame), even if this is not visible to the audience and the moving poster is designed in a loop. On the other hand, interactive posters allow the user to reshape the way that the information is presented. This rendering can be either direct, via user interaction, or indirect, via contextual ambient or user data. Only the interactive posters allow the incorporation of data through computational techniques, such as Computer Vision (CV) or Pattern Recognition. Therefore, they will remain perpetually in an unfinished state. In the context of this study, we will only focus on the development and study of this type of posters. Although this is a relatively new and unexplored area, some applications exist.

Briefly, interactive posters are systems that allow the user interaction. Studio Feixen's poster for *Oto Nové Swiss Festival*, at London's Cafe Oto (Studio Feixen & Giger, 2017), is a case study of this type of poster. In its digital version, this poster become in a web app which enables the user interaction, through the mouse. Although the set of interactions are predefined, the way the users interact with the poster is always different. Furthermore, Studio Feixen recent work, in this medium, is exploring the development of moving posters in both perspectives, either interactive (e.g. the *Sunset* poster (Studio Feixen, n. d.)) or animated (e.g. *Vlow!* identity (Studio Feixen, 2016)). Additionally, their projects have been well-accepted by the design community, achieving a high impact and exposure in specialised media, and increasing the design interest in this medium (see (Bourton, 2017) (Vesnin, 2017)).

On the other hand, the system can gather information from the environment without a user direct interaction. The campaign to Swedish pharmacy chain *Apotek*, in Stockholm subway, developed by Åkestam Holst, is a good case in point (Xie, 2014; Åkestam Holst, n. d.). The poster responds to the incoming trains, using the data recorded by ultrasonic sensors, ensuring, therefore, that the model has her hair tousled by the 'wind' of the moving train.

At the same time, other designers are investing the development of "poster machines" that enable the user to create a poster, through a previously developed framework (Armstrong & Stojmirovic, 2011). In a first stage, the system presents a default poster, and the user, through the manipulation of the system can see the modifications and create a poster design. Accordingly, these systems are interactive posters, according to our point of view.

In this sense, Luiz Ludwig developed an interactive user-oriented poster compositor, the *Poster Machine,* using an Arduino and Processing (Pelson, Kim, Dlugash, & Zotter; Ludwing, 2013). The user can create a unique poster design, shaping the scale and the position of the elements manipulating a series of knobs and switches. In the same perspective, Project Projects studio, in collaboration with the Kounkuey Design Initiative, developed the installation *Productive Posters* (Project Projects, 2008). This installation allows the user to write or draw in blank forms in order to develop a poster design. For each poster there are modules that can be included, for instance, to add quotations or give background information (Armstrong & Stojmirovic, 2011).

The use of CV techniques (such as motion capture or face recognition) to produce graphic images, it is not a new practice (Levin, 2006). In recent times, "data-driven design" artefacts start to use them in graphics context (e.g. (The Partners, 2017)) (Laranjo, 2017). Nevertheless, their use is still uncommon in this scenario. This is largely due to the advanced understanding required to the implementation of some CV algorithms which has been frightening graphic designers. However, nowadays a number widely-used and effective techniques can be implemented by everyone, and several toolkits offer an easy access to more advanced CV functionalities (Levin, 2006). The rapid increase of technological power and the necessity of 'moving posters' began to react to environmental stimuli, in our opinion, can lead these technologies to take a main role in the design process of this type of artefacts (Benyon, 2016).

The *Camera Postura,* by LUST, (LUST, 2014) can be recognised as one of the most innovative artefacts in this joint-venture between this two fields. This installation applies computer vision techniques to develop an interactive movie poster generator. The installation tracks the user's body language and attempts to match it with similar movie scene's poses. To the created images additional information (e.g. movie information) is added. Each pose results in the creation of a film poster at each interaction. This installation was implemented in the Netherlands Film Festival (2014) and the visitor's gestures matched with scenes of the 20 most popular films in the festival.

As above mentioned this is a field in fast changing. Other examples can be found in specialized websites and literature. While some effort has been made to provide references from books or papers, unfortunately, these projects are developed in recent times and the information is only available online.

### 3. The Approach

In this early stage, we start to develop a poster design that enables the variance of the position and the perception of its shapes. In order to make this variance possible, we intercalate three-dimensional geometries with two-dimensional shapes, the geometries are created in the 3D space, however the user only has this perception if he/she interacts with the poster. This first artefact is presented only as concept proof to study the viability of the project. The poster design is created in an interactive scenario using Processing (see Fig. 1). During this process, we also use the standard OpenCV Processing library, developed by Greg Borenstein, and Temboo software toolkit.

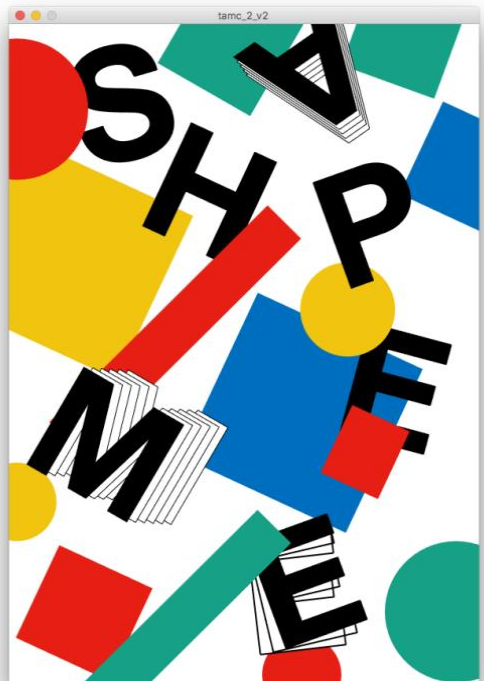

Figure 1: interactive poster design in its initial status. Demo videos are available at
https://student.dei.uc.pt/~srebelo/public/moving_poster/demo/

The poster in Fig. 1 is composed by 3 types of graphic elements: (1) boxes; (2) spheres; and (3) letters. The colours palette associated with poster changes in accordance with the temperature of the venue where the poster is installed. Furthermore, the user can select the element to reshape, by clicking on it. In this sense, the system gathers data from its surrounding environment in two ways: (1) directly, via user motion tracking; and (2) indirectly, using contextual ambient data, in this case, weather data.

  a.  *Motion Tracking*

The user motion is gathered by tracking the user's head, using a face detection algorithm. The position of the viewer's head is rebounded, in relation to the camera range, i.e. if the user has the head in the centre of the captured image, the poster does not change, however **the more the user** moves horizontally or/and vertically, the more noticeable is the change (see Figs. 2 and 3).

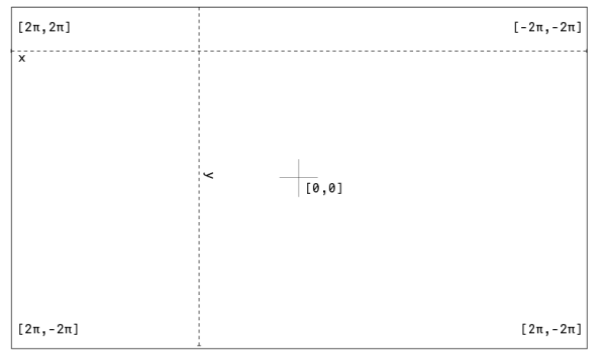

Figure 2: Motion capture mapping schema. The motion is mapped in a 2D coordinate, in radians, relative to the position of the head with the camera

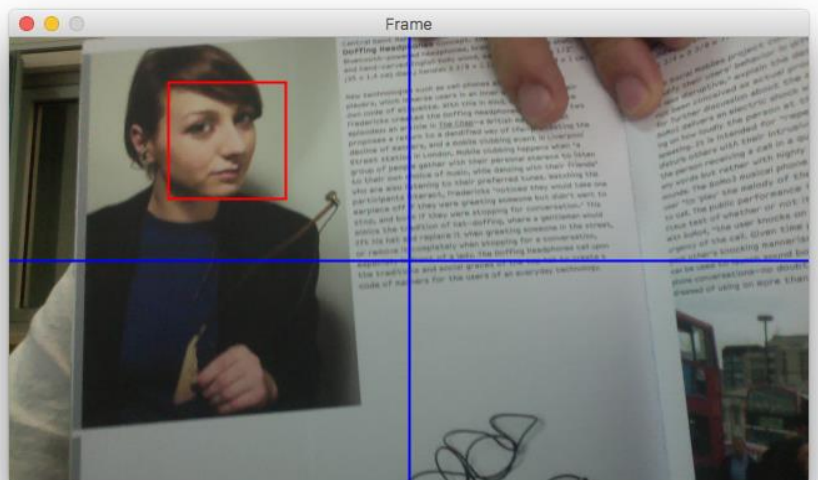

Figure 3: Motion capture method in use.

If more than one person is in front of the poster, the motion capture method calculates the average point between the positions of the viewers and uses this position.

*b. Contextual Ambient Data*

In this stage of the project, the system is only able to get weather data from the venue where the poster is installed. The information is used to assign a colour palette to the poster. This data is tracked by using the public Internet Protocol (IP) Address of the system to determine the place where the poster is connected to the Internet. Subsequently, the venue's information is used to get the current temperature of this place. The system updates the weather information at each five minutes.

Afterwards, the current temperature of the poster's venue is normalised by taking into consideration the average minimum and maximum temperature values in the poster location. In this case, we normalised the value taking in consideration the data about the climatological normal of air temperature provided by The Portuguese Institute for Sea and Atmosphere, for the city of Coimbra, in the period between 1981–2010. Accordingly, the minimum value is -4,5º C and the maximum 41,6º C (IPMA, 2011).

The different poster colours are added to the system in the form of a spectrum divided into five control points (see Fig. 4). The current colour value is then determined by placing the normalised temperature value in the colour spectrum. The colour is generated through the interpolation between the colour of the two closer key points.

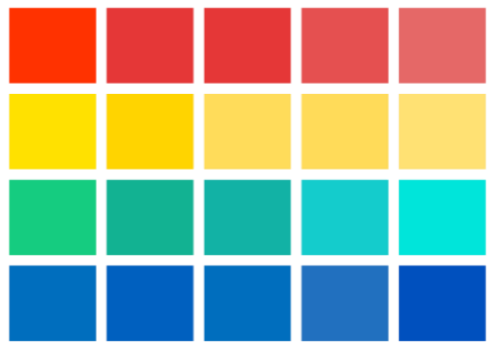

Figure 4: Control points of the colour spectrums added to the poster system.

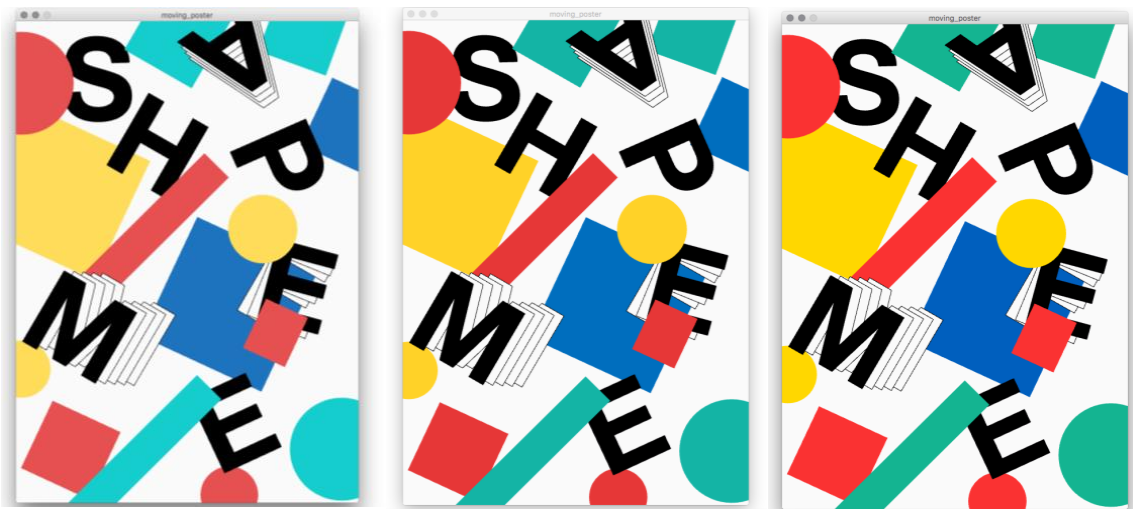

Figure 5: Posters designs rendered during times with different temperatures. From poster rendered in the coolest moment (in the left) to the rendered warmers periods.

## 4. Elements' Behaviour Design and Experimental Setup

As stated before, the poster is composed of three types of essential graphical elements (spheres, boxes and letters). Each element has a distinct behaviour.

   *a. Sphere Behaviour*

When the sphere is selected, it moves in one of the axes, i.e. the centre of spheres changes in the direction of *x* or *y*-axes. The translate axis is selected randomly in the initialisation of the object and, in the maximum, the spheres moves ten pixels in each direction. The user's motion is normalised using the cosine, in the *x*-axis, and the sine, in the *y*-axis (see fig. 7).

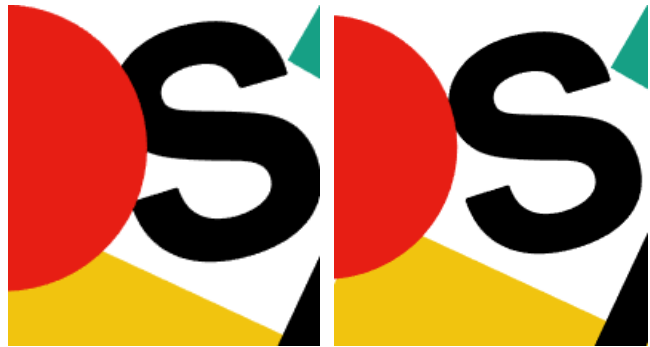

Figure 7: An example of the spheres' behaviour. On the left: The system does not recognise a user motion. On the right: the system employs a graphical translation to respond to a user's motion normalized in the value of -2π, i.e., the user has the head on the top right corner of the area captured by the camera.

   *b. Box Behaviour*

The boxes elements have a similar behaviour to the spheres, however, the translation is replaced by a rotation. As in spheres, the rotation axis is chosen randomly in the initialisation of the object. In this kind of elements, directional lights are used, creating shadows on the sides of the box and emphasising to the rotation (see Figs. 8 and 9).

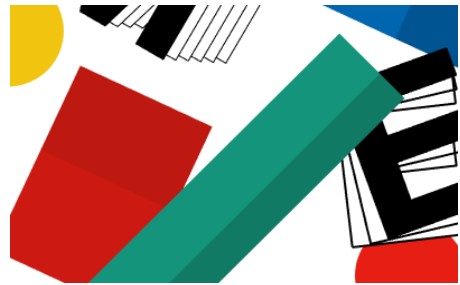

Figure 8: Example of a rotation applied to a box element. In the green box, the rotation is along the *x*-axis. In the red box, the rotation is along the *y*-axis.

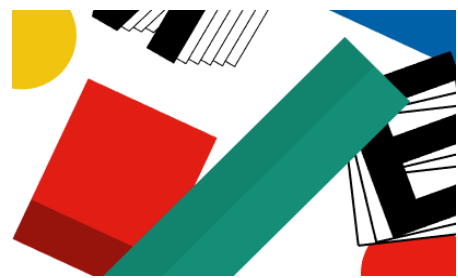

Figure 9: In the same scenario of Figure 7, rotation applied to a box with an inverse user's motion value.

   *c. Letters Behaviour*

The letters shapes have more dynamic and variable interactions. There are three types of interactions on letters: (1) rotation, i.e. the letter rotates around its centre; (2) rotation with dragging, i.e. the letter rotates around their upper right corner and leave a dragging effect; and, finally, (3) translate with dragging, i.e. the letter changes its position employing an accordion effect in its dragging. Additionally, the rotation with dragging has also two modes: (1) fixed, when the dragging is not created; and (2) non-fixed, when a new dragging is created.

In this prototype, the rotation effect is applied to the letters S and P. The rotation with dragging is applied in the E and H glyphs (the fixed mode in the E from word "me" and the non-fixed in the other letters). The translate with dragging is used in the remaining letters.

The rotation effect is similar to the one applied to other poster elements (see Fig. 10). The rotation with dragging effect applies a rotation to letter in its upper right corner, creating drags. This effect has two modes. The first mode, the non-fixed, creates drags when the rotation is done (see Fig. 11). Alternatively, the fixed mode does not create new drags and rotate also the existing drags (see Fig. 12). These number of drags are created whenever the rotation angle is a multiple of the initial angle rotation. The direction of rotation is randomly selected in the initialisation.

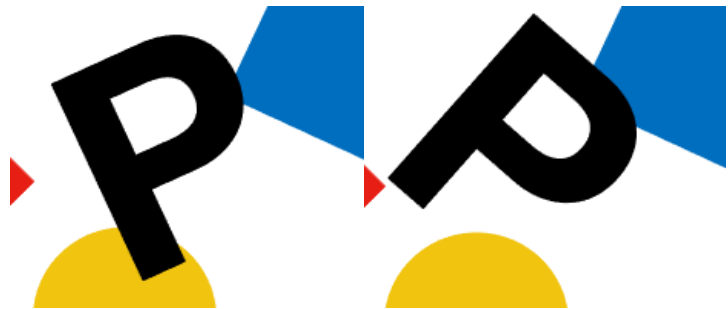

Figure 10: Example of a 'rotation' effect in a letter.

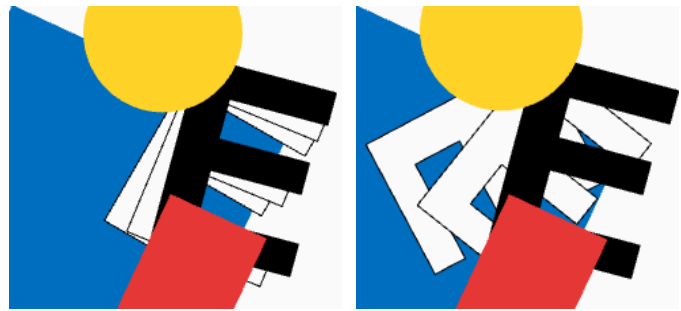

Figure 11: Example of a 'rotation with dragging' effect, in its fixed mode.

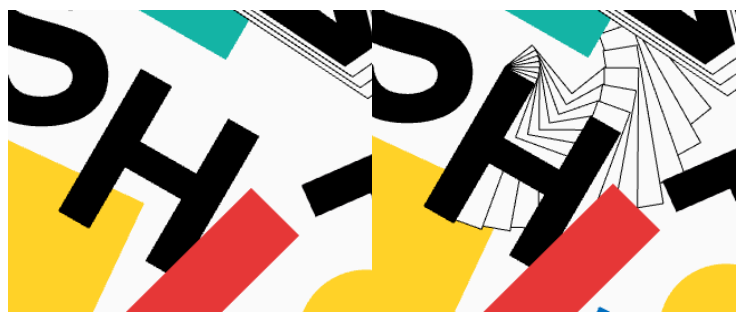

Figure 12: Example of a 'rotation with dragging' effect, in its non-fixed mode.

In the translate with dragging effect, the glyph is translated, according to the user motion value, and, consequently, the space between the drags increases or decreases (see Fig. 13).

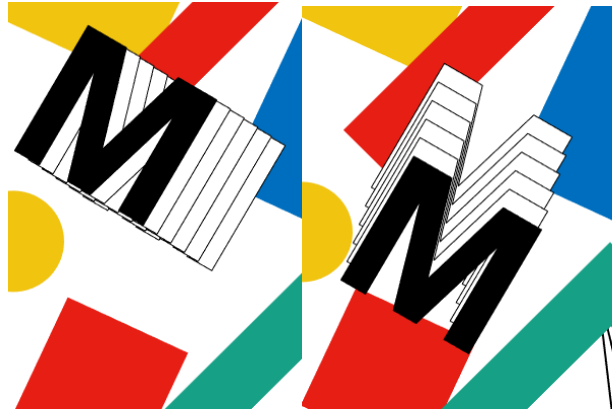

Figure 13: Example of a 'translate with dragging' effect. On the left: the initial stage of the poster; On the right: The effect applied to a glyph.

## 5. Conclusion

We have presented an experiment to develop a 'moving poster' that integrates external data in its process of shaping. Although this is a work in process, the system already translates environmental data in graphic information. The data is gathered directly from the user, tracking the viewer's position, and, indirectly, via contextual ambient data. This allows us to create different user experiences and to develop at each interaction a unique poster design. The user motion data is translated into graphical information using computer vision techniques, namely face detection.

Although this experimentation serves mainly as a proof of concept, the present results motivate us to further explore this idea. With only a small number of parameters (i.e. the number of people that watch the poster, their position, and the temperature of the venue), the dynamism of the poster is guaranteed and the graphical change is significant. In another point of view, we proved with this experiment that the absence of synergetic work between these fields is not directly related to the difficulty in connecting them. In this way, we believe that this opened a window of opportunity to create novel and experimental work. In future iterations of the project, we expect to explore more advanced motion tracking techniques, image processing, and automatic poster composition.

Future work will also focus on: (1) increasing the number of features tracked from the user and the environment; (2) developing an automatic poster composition system; (3) increasing the number of transformation methods in the poster; and (4) developing a physical implementation of the system.